%% file: radcor.tex
\documentclass[12pt]{article}
\usepackage{epsfig,french,rotating}
 
\newcommand\pubnumber{
}
\newcommand\pubdate{
}
\newcommand\hepnumber{
}
 
\def\csumb{{Université de Haute Alsace\\
 Mulhouse, France}}
 
\def\Title#1{\begin{center} {\Large\bf #1 } \end{center}}
\def\Author#1{\begin{center}{ \sc #1} \end{center}}
\def\Address#1{\begin{center}{ \it #1} \end{center}}

\newcommand\pubblock{\rightline{\begin{tabular}{l} \pubnumber\\
         \pubdate\\ \hepnumber \end{tabular}}}
\newenvironment{Abstract}{\begin{quotation}  }{\end{quotation}}
\newenvironment{Presented}{\begin{quotation} \begin{center}
             Presented at the\end{center}
      \begin{center}\begin{large}}{\end{large}\end{center} \end{quotation}}

\makeatletter
\def\section{\@startsection{section}{0}{\z@}{5.5ex plus .5ex minus
 1.5ex}{2.3ex plus .2ex}{\large\bf}}
\def\subsection{\@startsection{subsection}{1}{\z@}{3.5ex plus .5ex minus
 1.5ex}{1.3ex plus .2ex}{\normalsize\bf}}
\def\subsubsection{\@startsection{subsubsection}{2}{\z@}{-3.5ex plus
-1ex minus  -.2ex}{2.3ex plus .2ex}{\normalsize\sl}}
 
\renewcommand{\@makecaption}[2]{%
   \vskip 10pt
   \setbox\@tempboxa\hbox{\small #1: #2}
   \ifdim \wd\@tempboxa >\hsize     
       \small #1: #2\par          
     \else                        
       \hbox to\hsize{\hfil\box\@tempboxa\hfil}
   \fi}
 
 \def\citenum#1{{\def\@cite##1##2{##1}\cite{#1}}}
 
\newcount\@tempcntc
\def\@citex[#1]#2{\if@filesw\immediate\write\@auxout{\string\citation{#2}}\fi
  \@tempcnta\z@\@tempcntb\m@ne\def\@citea{}\@cite{\@for\@citeb:=#2\do
    {\@ifundefined
       {b@\@citeb}{\@citeo\@tempcntb\m@ne\@citea\def\@citea{,}{\bf ?}\@warning
       {Citation `\@citeb' on page \thepage \space undefined}}%
    {\setbox\z@\hbox{\global\@tempcntc0\csname b@\@citeb\endcsname\relax}%
     \ifnum\@tempcntc=\z@ \@citeo\@tempcntb\m@ne
       \@citea\def\@citea{,}\hbox{\csname b@\@citeb\endcsname}%
     \else
      \advance\@tempcntb\@ne
      \ifnum\@tempcntb=\@tempcntc
      \else\advance\@tempcntb\m@ne\@citeo
      \@tempcnta\@tempcntc\@tempcntb\@tempcntc\fi\fi}}\@citeo}{#1}}
\def\@citeo{\ifnum\@tempcnta>\@tempcntb\else\@citea\def\@citea{,}%
  \ifnum\@tempcnta=\@tempcntb\the\@tempcnta\else
  {\advance\@tempcnta\@ne\ifnum\@tempcnta=\@tempcntb \else\def\@citea{--}\fi
    \advance\@tempcnta\m@ne\the\@tempcnta\@citea\the\@tempcntb}\fi\fi}
\makeatother

\input econfmacros2.tex
 
\begin{document}
\begin{titlepage}
\pubblock
 
\vfill
\def\thefootnote{\fnsymbol{footnote}}
\Title{Higgs and Supersymmetry searches at the Large Hadron Collider}
\vfill
\Author{Fran\c{c}ois Charles}
\Address{\csumb}
\vfill
\begin{Abstract}
We present here the results for Higgs and Supersymmetry prospective
searches at the Large Hadron Collider. We show that for one year
at high luminosity, Standard Model and MSSM Higgs should be observed
within the theoretically expected mass range. MSUGRA and restricted
phenomenological MSSM searches lead to discovery of up to $2.5$ $TeV$
squarks and gluinos.  
\end{Abstract}
\vfill
\begin{Presented}
5th International Symposium on Radiative Corrections \\
(RADCOR--2000) \\[4pt]
Carmel CA, USA, 11--15 September, 2000
\end{Presented}
\vfill
\end{titlepage}
\def\thefootnote{\arabic{footnote}}
\setcounter{footnote}{0}
\section{Higgs searches}
\subsection{Standard Model Higgs}
\begin{figure}[h!]
	\begin{center}

\hspace{-9.8cm}
\begin{rotate}{-90}          
	\includegraphics[width=7cm,draft=false]{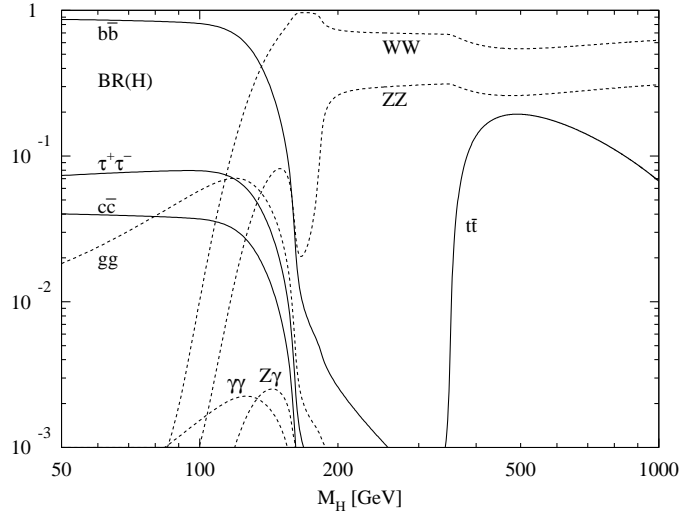}
\end{rotate}          
	\end{center}

\vspace{7.cm}
	\caption{Branching ratios for major Higgs decay channels as a function
	of its mass.}
	\label{h_br}
	\end{figure}

The search for the Higgs boson is one of the major task of the LHC.
Several decay channels of Higgs have been explored at the LHC among them: ZZ,WW and $\gamma \gamma$
provide the best discovery possibilities. We can see in figure \ref{h_br} the expected
branching ratio as function of the Higgs mass. We can notice that $b\bar{b}$
and WW are the dominant mode.
 
	\begin{figure}[hbt]
	\begin{center}
	\includegraphics[bb=-78 180 630 630,width=8cm,clip=true,draft=false]{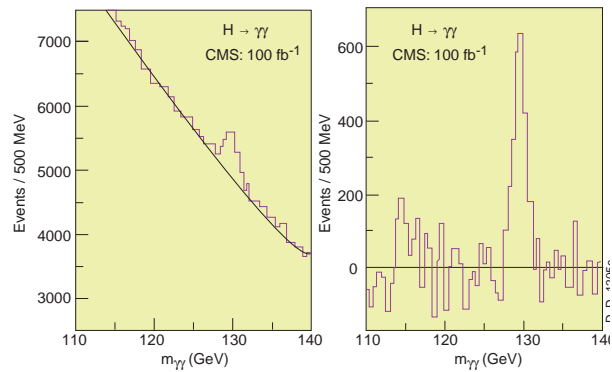}
	\end{center}
	\caption{$ H \rightarrow \gamma \gamma$ with CMS}
	\label{h_ga}
	\end{figure}

One of the most important and challenging mode is $ H \rightarrow \gamma \gamma$
in the Higgs mass range 100 GeV to 150 GeV. It requires excellent calorimetric precision
both in energy and angular measurement. True $\gamma \gamma$ production as well as reducible
$\gamma jet$ ( with $\pi^0$ faking photon) should constitute the main
background.
The figure \ref{h_ga} illustrate the diphoton invariant mass as expected with CMS detector
for one year at high luminosity.
 
\begin{figure}[hbt]
	\begin{center}
\epsfig{file=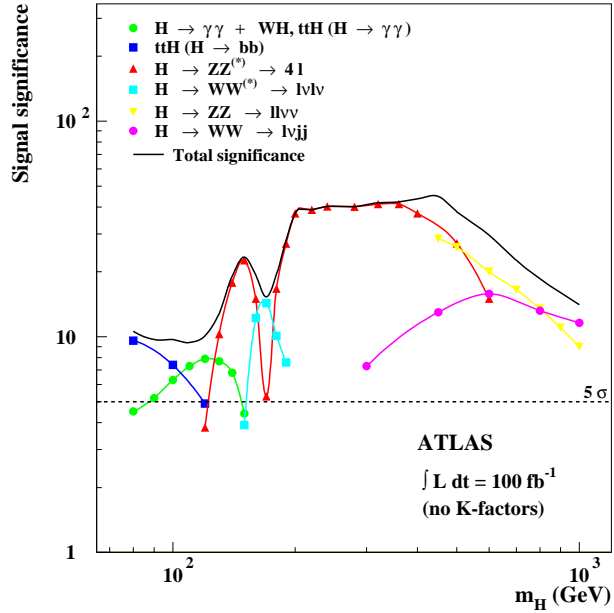,width=8cm}
	\end{center}
	\caption{Higgs boson discovery with ATLAS, high luminosity}
	\label{sign_l}
	\end{figure}

We can see in figure \ref{sign_l} the expected significance ($S/\sqrt{B}$) for one year of high
luminosity in ATLAS \cite{ATLAS} experiment over the full Higgs mass
range. Combining all possible
channels should give more than 5 $\sigma$,
even in the low luminosity regime.
 
\subsection{Supersymmetry Higgses}
\begin{figure}[hbt]
	\begin{center}
	\epsfig{file=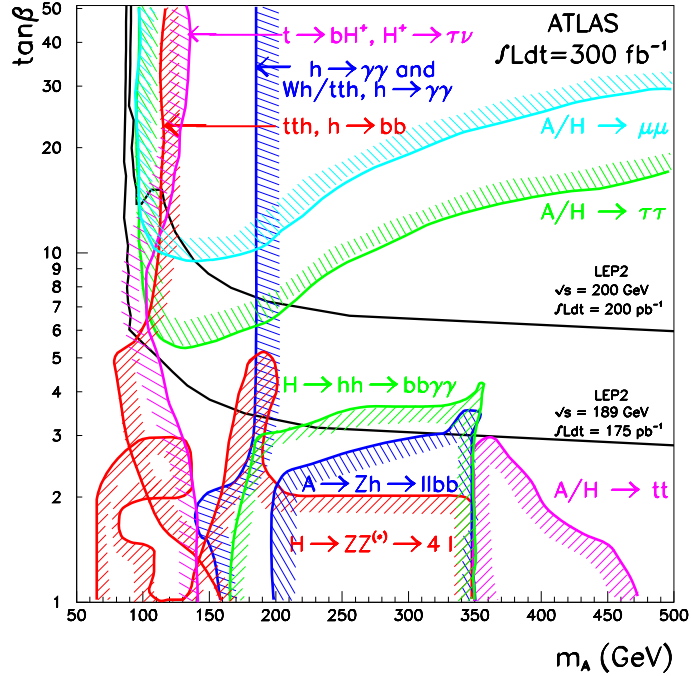,width=9cm}
	\end{center}
	\caption{MSSM Higgs boson discovery with ATLAS, high luminosity}
	\label{h_mssm}
	\end{figure}

In the Minimal Supersymmetric Standard Model, we expect 5 Higgses, two charged
and 3 neutral Higgses. For the lightest	Higgs, $h\rightarrow b b$ remains the most important
mode while for heavier Higgs: $H/A \rightarrow \tau \tau,tt,\mu \mu $ are interesting modes.
As can be seen from figure \ref{h_mssm} in the case of minimal mixing, the full range $m_A=50-500$ $GeV$
,$tan\beta=1-50$ can be covered. The most difficult region correspond to intermediate $m_A $ and
$tan\beta$. In the case of light SUSY particles we would also observe decay of Higgs into a pair of
sparticles.

\section{Supersymmetry searches}
 
\subsection{Introduction}
 
One of the aims of the LHC collider is to search for the physics
beyond the Standard Model (SM). One of the direction of this search is
a possible discovery of  superpartners of ordinary particles as expected in
Supersymmetric extensions of SM (SUSY).
SUSY, if it exists, is expected to reveal itself at LHC via excess of
 (multilepton +) multijet + E$_{T}^{miss}$ final states
 compared to Standard Model (SM) expectations .
 
The main goal of the LHC studies is to evaluate the potential of the CMS and ATLAS detectors
\cite{CMS,ATLAS}, to find evidence for SUSY. It deals, first with a semi-inclusive search
mainly squarks and
gluino mass reach, as the production cross section of these strongly
interacting sparticles (pair production or in association with charginos and
neutralinos) dominates the total
SUSY cross section over a wide region of the parameter space.
These studies were done first in the framework of MSUGRA model and at a later stage
in a constrained MSSM.
 
 Finally we will consider more exotic models
like R-parity violated and Gauge Mediated Supersymmmetry Breaking models.
 
\subsection{MSUGRA Studies}
 
The large number of SUSY parameters even in the framework of
Minimal extension of the SM (MSSM) makes it difficult to evaluate the
general reach. So, for this study
we restrict ourselves  at present to the mSUGRA-MSSM model. This model
evolves from MSSM, using Grand Unification Theory (GUT) assumptions
(see more details in e.g. \cite{gdr}).
In fact, it is a representative model, especially in case of inclusive studies
and reach limits are expressed in terms of squark and gluino mass
values.
 
The mSUGRA model contains only five free parameters $:$
\begin{itemize}
 \item a common gaugino mass ($m_{1/2}$) ;
 \item a common scalar mass ($m_{0}$);
 \item a common trilinear interaction amongst the scalars ($A_{0}$);
 \item the ratio of the vacuum expectation values of the Higgs fields;
 \item a Higgsino mixing parameter $\mu$ which enters only through its
   sign ($sign (\mu)$).
\end{itemize}
 
\subsubsection{Simulation procedure}
 
The PYTHIA 5.7 generator \cite{PYTHIA} is used to generate  all
 SM background processes, whereas ISAJET 7.32 \cite{isajet}
is used for mSUGRA signal simulations.
    The CMSJET (version 4.51) fast MC package \cite{CMSJET}
 is used to model the CMS detector response, since it still looks
impossible to perform a full-GEANT \cite{geant} simulation for the present study,
 requiring to
process multi-million samples of signal and SM background events.
 
The SM background processes considered are $:$  QCD  2 $\rightarrow$ 2
 (including $b\bar{b}$), $t\bar{t}$, $W+jets$, $Z+jets$. The $\hat{p}_{T}$
range of all the background processes is subdivided into several  intervals to
facilitate  accumulation of statistics in the high-$\hat{p}_{T}$ range $:$
100-200 GeV, 200-400 GeV, 400-800 GeV and  $>$ 800 GeV (additional interval
of 800-1200 GeV is reserved for QCD).
The accumulated SM background statistics for all background channels
amounts to about 200 millions events.  
 
The grid of probed m$_{0}$, m$_{1/2}$ mSUGRA points has a cell size of
 $\Delta$m$_{0}$=$\Delta$m$_{1/2}$=100 GeV for m$_{0}<$1000 GeV and
 $\Delta$m$_{0}$=200 GeV,
 $\Delta$m$_{1/2}$=100 GeV for  m$_{0}>$1000 GeV.
This was also probed with the appropriate mixture of signal and pile-up
events.
The kinematics of signal events
is usually  harder than that of SM background
for the interesting regions of maximal reach of squark-gluino masses .
The cross section of the background is however higher by orders of magnitude
and high-p$_{T}$ tails of different backgrounds
can have a kinematics similar to that of the signal.
 
\begin{figure}[hbtp]
\begin{center}
\vspace*{0mm}
\epsfig{file=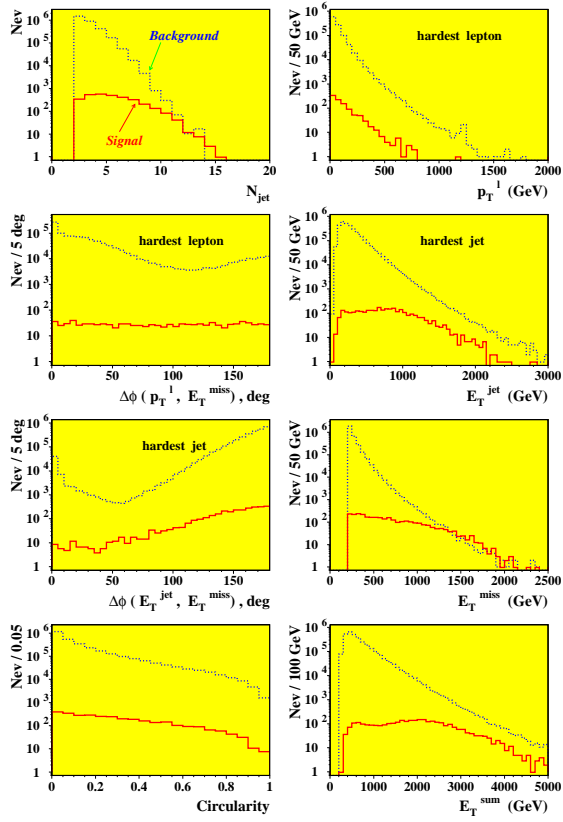,width=8cm}
\caption{ Comparison of the mSUGRA signal and SM background,
 with 100 fb$^{-1}$,
in one point of mSUGRA parameter space $:$ m$_0$=1000 GeV, m$_{1/2}$=800 GeV
(m$_{\tilde{g}}$$\approx$m$_{\tilde{q}_{L}}$$\approx$ 1900 GeV)
for the E$_T^{miss}$ signature. Initial cuts are applied.}
\label{comp}
\end{center}
\end{figure}

In figure \ref{comp}
 we compare some signal distributions for the  point
 (m$_{0}$, m$_{1/2}$) = (1000,800),
 corresponding to  m$_{\tilde{g}} \approx$ m$_{\tilde{q}_{L}} \approx$
 1900 GeV,
m$_{\tilde{\chi}_{1}^{0}}$ = 351 GeV, m$_{\tilde{\chi}_{2}^{0}}$ =
 m$_{\tilde{\chi}_{1}^{\pm}}$ = 668 GeV,
and distributions of the sum of all SM background processes 
for the E$_{T}^{miss}$  signature.

Both signal and
background histograms contain only events satisfying first level selection
criteria ($N_{jets}>2$ and $E^{miss}_T>200GeV$).
 Only the hardest jet and lepton in the event are shown
 in distributions in figure \ref{comp}.
 
Since the topology of signal and background events is rather
similar already  after  first level selection cuts,
the difference in the
angular distributions and circularity is not significant either,
it is thus not very useful to apply
cuts on these variables too. The difference in the lepton p$_{T}$ distributions
 is also not very pronounced as  signal leptons are produced in
cascade decays, thus loosing  ``memory'' about the hardness of the original
process. But for extremely high masses of squarks or gluinos ($\sim$ 2 TeV),
there is some difference in the angular and p$_T^{l}$ distributions
between signal and the total SM background.
One can deduce from figures \ref{comp} that cuts on the jet multiplicity N$_{j}$
 and E$_{T}^{miss}$
are the most profitable ones for background suppression .
Of course, there is inevitable correlation between variables both in signal
and background, e.g. an obvious correlation  between
 E$_{T}^{miss}$ and the hardest jet  E$_{T}$ in QCD events, since there
 E$_{T}^{miss}$ is mainly produced by neutrinos from b-jets and/or
 high-E$_{T}^{jet}$ mis-measurement. This can lead to a degradation of
the efficiency of some cuts, if fixed cuts are used.
It is thus more profitable to have
 adjustable cuts to meet various kinematical conditions
in various domains of mSUGRA parameter space and take into account
difference in topology between various signatures. For this reason,
we search among 10000 cuts combination, the best selection at each generated
MSUGRA point. 

\subsubsection{Results}

\begin{figure}[hbtp]
\begin{center}
\vspace*{0mm}
\hspace*{0mm}\resizebox{0.99\textwidth}{19cm}
      {\includegraphics{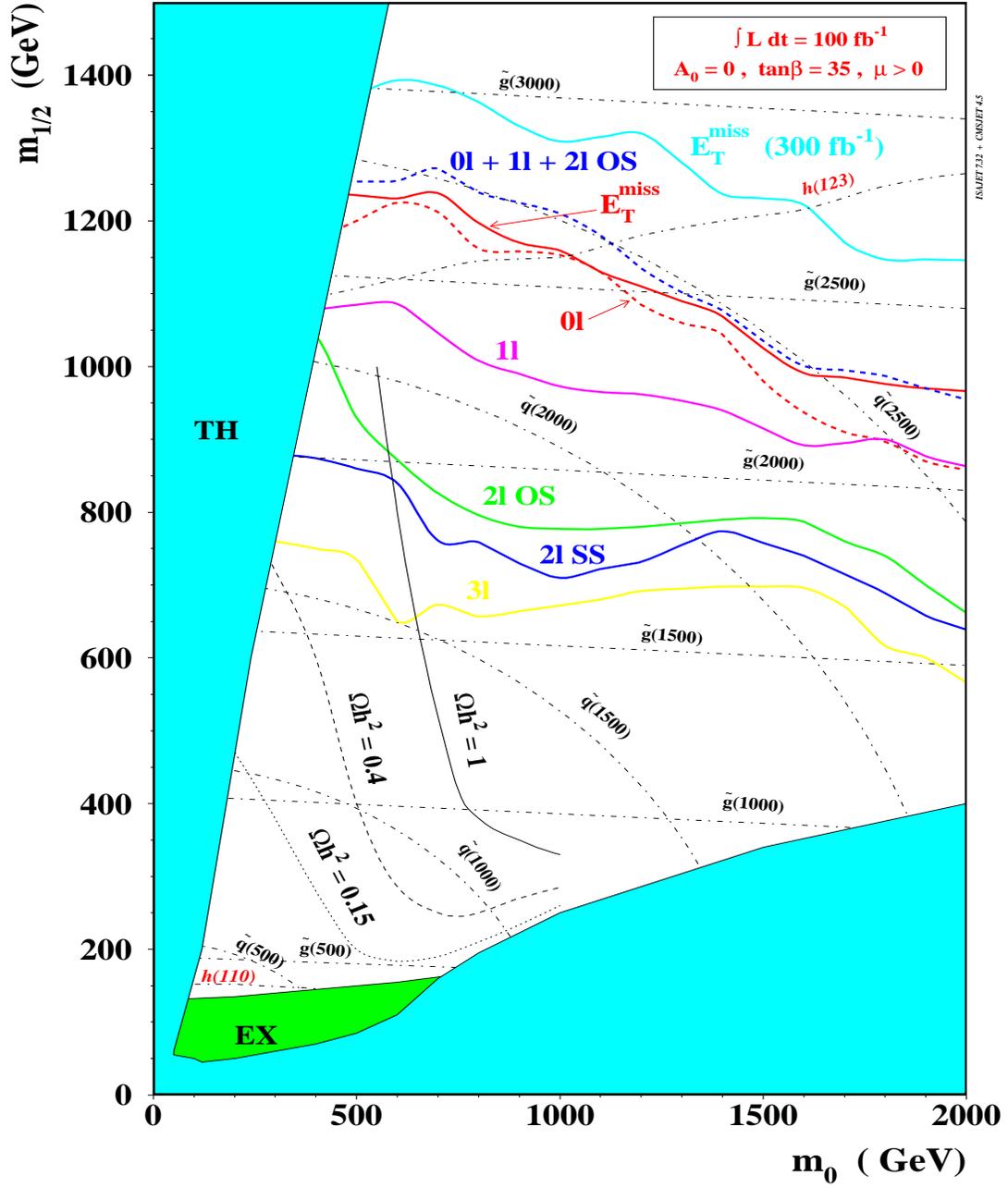}}
\caption{ 5 sigma reach contours for various final states
with 100 fb$^{-1}$ (see also comments in text).}
\label{c35+}
\end{center}
\end{figure}

\begin{figure}[hbtp]
\begin{center}
\vspace*{0mm}
\hspace*{0mm}\resizebox{0.99\textwidth}{20cm}
        {\includegraphics{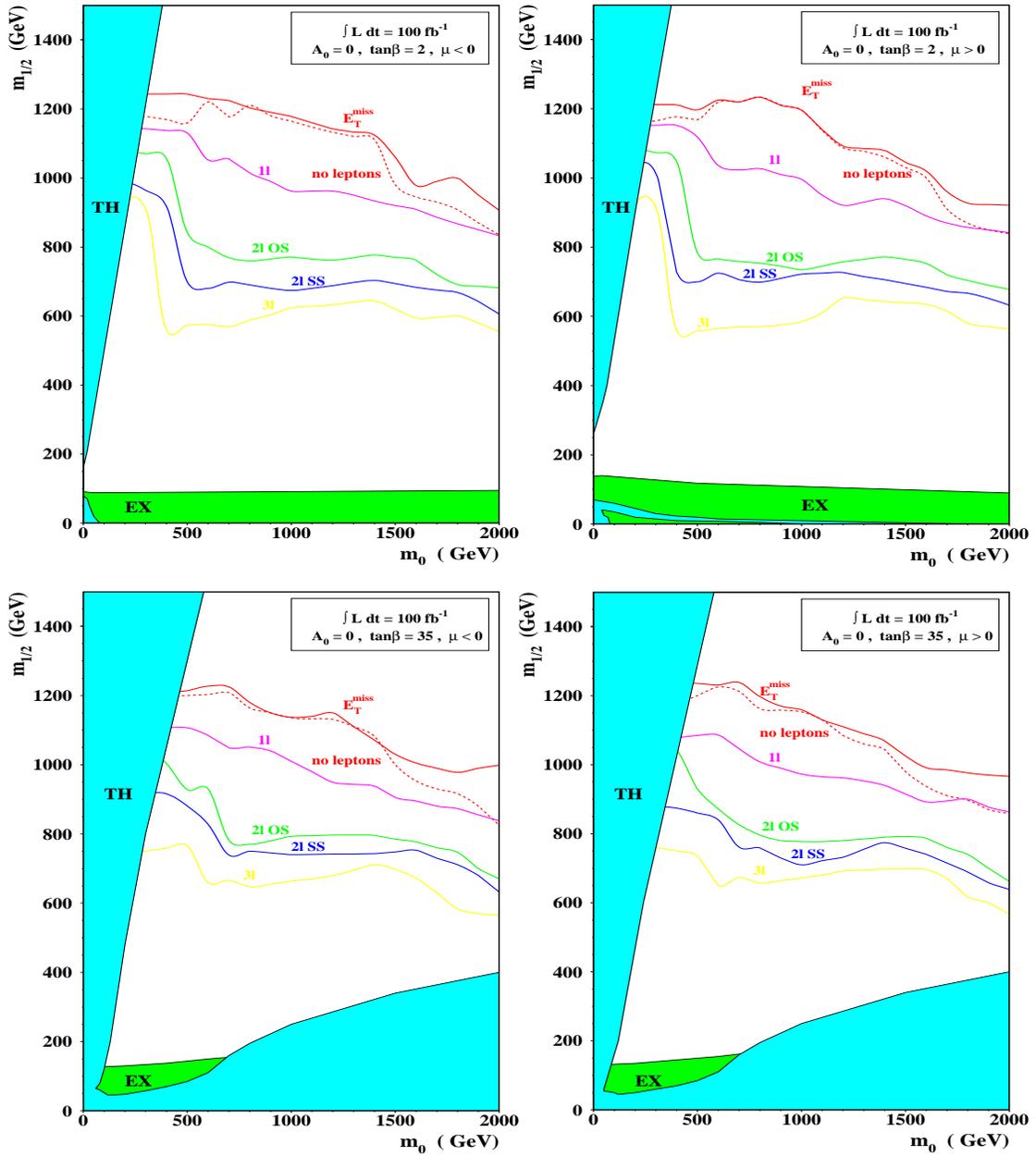}}
 \caption{ Simplified figures gathered together.}
\label{comb}
\end{center}
\end{figure}

Figures \ref{c35+} and \ref{comb} show the main results of our study for
mSUGRA  assuming an integrated luminosity of 100 fb$^{-1}$.
Figure \ref{c35+} contains isomass contours for
squarks ($\tilde{q}$), gluino ($\tilde{g}$)  and lightest scalar Higgs
 ($h$). Numbers in parenthesis denote mass values of corresponding isomass
contour. The neutralino relic density contours
from ref. \cite{ex_regions}, for mSUGRA domain  m$_{0}$$<$1000 GeV,
 m$_{1/2}$$<$1000 GeV, are  shown 
for $\Omega h^{2}$ = 0.15, 0.4 and 1.0.
Value  $\Omega h^{2}$$>$1 would lead to a Universe age less than 10
billion years old, in contradiction with estimated age of the oldest stars.
The region in between 0.15 and 0.4 is favoured by the Mixed Dark Matter (MDM)
cosmological models. It is  a rather general situation that for all investigated sets of
mSUGRA parameters the best reach can be obtained with the E$_{T}^{miss}$
 signature. The more leptons required - the smaller reach, as can be seen from
figures \ref{comb}.
 The cosmologically
preferred region  $\Omega h^{2}$$<$0.4 seems to be entirely within the reach
 of CMS. In 
figure \ref{c35+}
 we also show our calculations for  the E$_{T}^{miss}$
signature reach for an integrated luminosity of 300 fb$^{-1}$,
 trying to estimate
the ultimate CMS reach.
 
\newpage
\subsection{MSSM}

\subsubsection{A restricted pMSSM } 
 
The model used in the next study is a phenomenological MSSM (19 parameters),
but with a further reduction of the
number of the parameters which allows us to perform simpler simulation, while
keeping the diversity of the signatures of MSSM events . We take into account,
respectively, the mass unification of squarks and sleptons (universality of
the three particles generations) and the unification of trilinear coupling.
This leads to 9 free parameters: \( \tan \beta  \), \( M_{A} \), \( \mu  \),
\( M_{1} \), \( M_{2} \),\( M_{\widetilde{g}} \), \( M_{\widetilde{q}} \),
\( M_{\widetilde{l}} \), \( A_{3} \).

\subsubsection{Signal}
 
The goal of this study is to evaluate the capacity of CMS detector
to observe MSSM signals. Two reasons
urge us to study this pMSSM after the MSUGRA study. First , MSUGRA is a rather
constrained model with only five free parameters. Contrary to MSUGRA, the pMSSM
has no fixed hierarchy of masses as shown later. Moreover, we try to estimate
MSSM parameter values using kinematical quantities measured by CMS, using a
fast simulation program.
We use a model with 9 parameters, which constitute an hyperspace with 9 dimensions.
In order to facilitate the analysis, we use a discretization of the parameters.
The choice of the number of value of each parameters depends of the parameter
sensitivity. We used a grid for squarks and gluinos masses with 9 values evenly
spaced between 600 and 3000 GeV, because the events characteristics at LHC depend
primarily on these two masses. On the other hand, the observables are not very
dependant on the parameter \( \tan \beta  \) and thus we only use two values
in order to distinguish the larges ones and the small values from it parameters.
Thus, the values which we selected for each parameter of this analysis are :
 
\begin{itemize}
\item \( M_{\widetilde{l}} \) : \( 200,\, 1000,\, 3000 \) \( GeV \)
\item \( M_{1} \) : \( 100,\, 500,\, 1000,\, 2000 \)\( GeV \)
\item \( M_{2} \) : \( 100,\, 500,\, 1000,\, 2000 \)\( GeV \)
\item \( M_{A} \) : \( 200,\, 1000,\, 3000 \)\( GeV \)
\item \( A_{3} \) : \( 0,\, 2000 \) \( GeV \)
\item \( \mu  \) : \( 200,\, 500,\, 2000 \) \( GeV \)
\item \( \tan \beta  \) : \( 2,\, 50 \)
\item \( M_{\widetilde{q}} \) : \( 600,\, 900,\, 1200,\, 1500,\, 1800,\, 2100,\, 2400,\, 2700,\, 3000 \)
\( GeV \)
\item \( M_{\widetilde{g}} \) : \( 600,\, 900,\, 1200,\, 1500,\, 1800,\, 2100,\, 2400,\, 2700,\, 3000 \)
\( GeV \)
\end{itemize}
We end up with a total of 140000 different combination of parameters for each
of which we generate 1000 events, a compromise between the limits imposed by
handling the data flow and sufficiently small statistical errors. The theoretical
and experimental constraints make it possible to reduce the number of combination
to a total of \( 17.10^{3} \).
The background production was estimated using Standard Model events leading
to similar signatures as MSSM events, in order to study the possibility of extracting
SUSY signals from the background.
We illustrate in the following figure the background and signal
distribution for various variables.
 
\begin{figure}
{\centering \resizebox*{9cm}{9cm}{\includegraphics{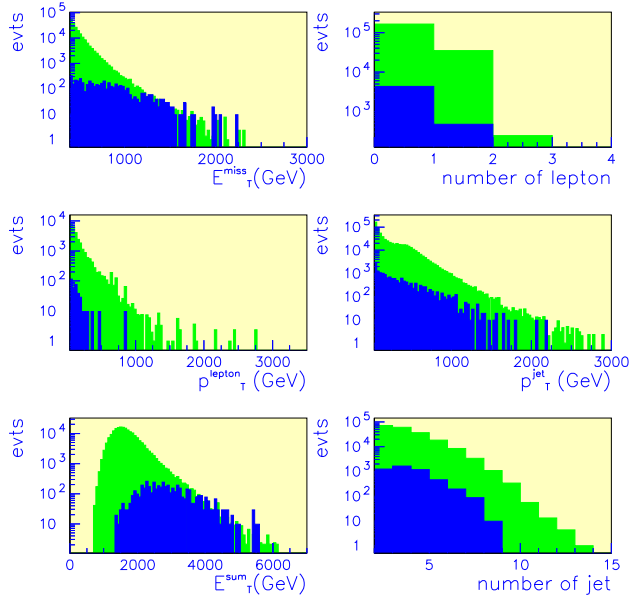}} \par}
\caption{Example 1: distribution of signal and background for differents observables
\label{fig: S/N 1}}
\end{figure}
The values of MSSM parameter for the figure \ref{fig: S/N 1} are : \\
\( M_{\widetilde{l}}=1000\, GeV, \) \( M_{1}=100\, GeV, \) \( M_{2}=500\, GeV, \)
\( M_{\widetilde{q}}=1800\, GeV, \)
\( M_{\widetilde{g}}=1800\, GeV, \) \( M_{A}=200\, GeV, \) \( \tan \beta =50, \)
\( \mu =2000\, GeV, \) \( A_{3}=2000\, GeV \). \\
Despite the corresponding low cross section \( \sigma =77fb \), the significance
of this parameter set is, after application of optimisation cuts, equal to \( 21 \).

 \subsubsection{case of large and close hierarchy of masses.} 
 
\begin{figure}
{\centering \resizebox*{9cm}{9cm}{\includegraphics{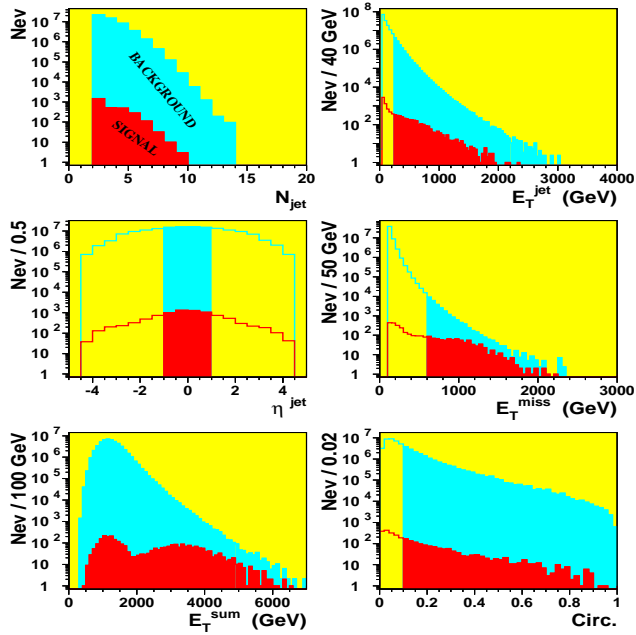}} \par}
\caption{Example 2: distribution of signal and background \label{fig: masse 1}}
\end{figure}

\begin{figure}
{\centering \resizebox*{10cm}{7cm}{\includegraphics{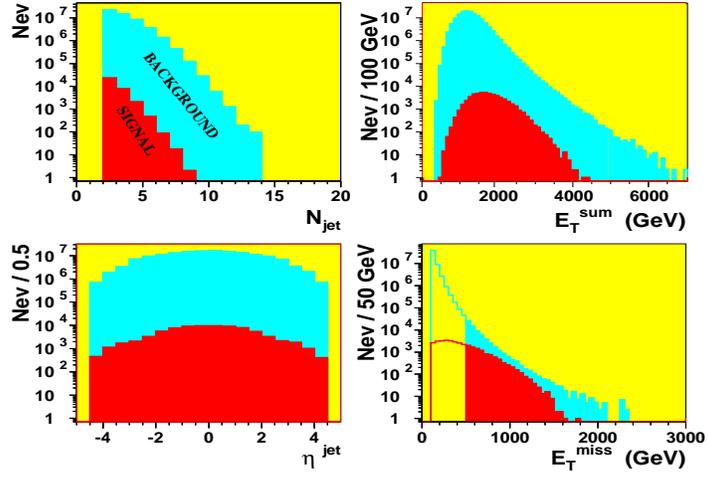}} \par}
\caption{Example 3: distribution of signal and background \label{fig: masse 2}}
\end{figure}

For the first example (figure \ref{fig: masse 1}) the masses of neutralinos
are much lower than the masses of squarks, gluinos and sleptons, the production
will be dominated by neutralinos and charginos. The parameter values are the
following : \\
\( M_{\widetilde{l}}=2000\, GeV, \) \( M_{1}=500\, GeV, \) \( M_{2}=500\, GeV, \)
\( M_{\widetilde{g}}=2000\, GeV, \) \( M_{\widetilde{q}}=2000\, GeV, \) \( M_{A}=1000\, GeV, \)
\( \tan \beta =50, \) \( \mu =200\, GeV, \) \( A_{3}=0\, GeV \). \\
The cross section of this set of parameters is \( \sigma =1.22pb \) and despite
the abundance of neutralinos, the low production rate of gluinos and squarks
allows nevertheless to obtain a significance equal to \( 10.2 \). The
cuts for figures \ref{fig: masse 1} and \ref{fig: masse 2},
are those giving the maximum significance (among the list of the
optimization cuts).
 
For the second example (figure \ref{fig: masse 2}) the parameter values are
the following: \( M_{\widetilde{l}}=1500\, GeV, \) \( M_{1}=940\, GeV, \)
\( M_{2}=2000\, GeV, \) \( M_{\widetilde{g}}=1000\, GeV, \)
\( M_{\widetilde{q}}=1020\, GeV, \) \( M_{A}=1000\, GeV, \) \( \tan \beta =50, \)
\( \mu =1050\, GeV, \) \( A_{3}=0\, GeV \).
The masses of neutralinos, gluinos, squarks and sleptons are comparable. The
main production proceeds via gluinos and squarks with a cross section \(
\sigma =2.0\, pb \)
and a significance equal to \( 36.3 \).
 
\begin{figure}
{\centering \resizebox*{10cm}{10cm}{\includegraphics{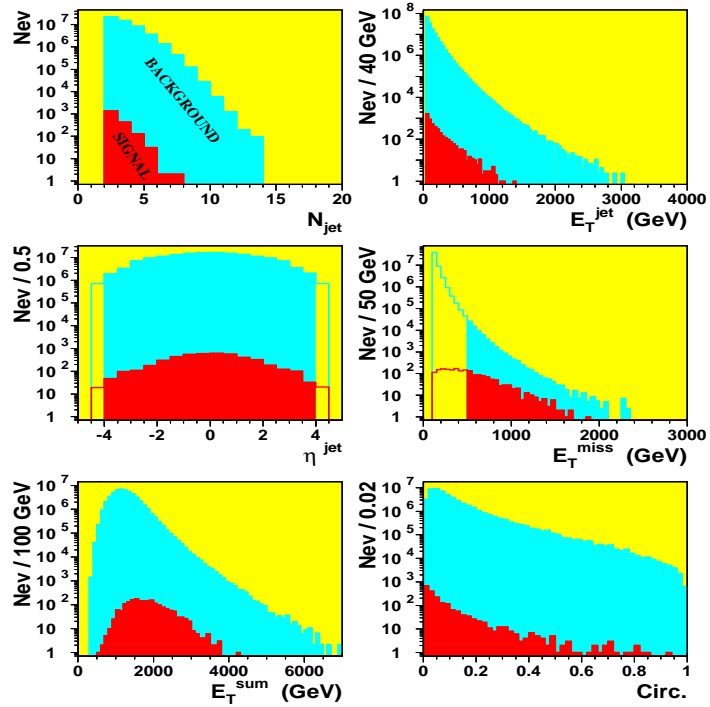}} \par}

\caption{Distribution of signal and background \label{fig: masse3}}
\end{figure}

For the third example (figure \ref{fig: masse3}) the parameter
values are the
following:
 
\( M_{\widetilde{l}}=1520\, GeV, \) \( M_{1}=1450\, GeV, \) \( M_{2}=2000\, GeV, \)
\( M_{\widetilde{g}}=1500\, GeV, \)
 
\( M_{\widetilde{q}}=1520\, GeV, \) \( M_{A}=1000\, GeV, \) \( \tan \beta =50, \)
\( \mu =1500\, GeV, \) \( A_{3}=0\, GeV \)
 
The masses of neutralinos, gluinos, squarks and sleptons are comparable but
with a higher value. The main production proceeds always via gluinos and
squarks
with, in this case, a cross section \( \sigma =0.126\, pb \) and a significance
equal to \( 3.2 \).
 
Even for the sets of parameters which would seem difficult (hierarchy of very
close mass or on the contrary very separeted), this method make it possible
to obtain good results, but we observe a limitation in the case of close
hierarchy
of masses with a dicovery limit of about 1.5 TeV instead of 2.5 TeV in the
other case.

\begin{figure}
{\centering \resizebox*{8cm}{8cm}{\includegraphics{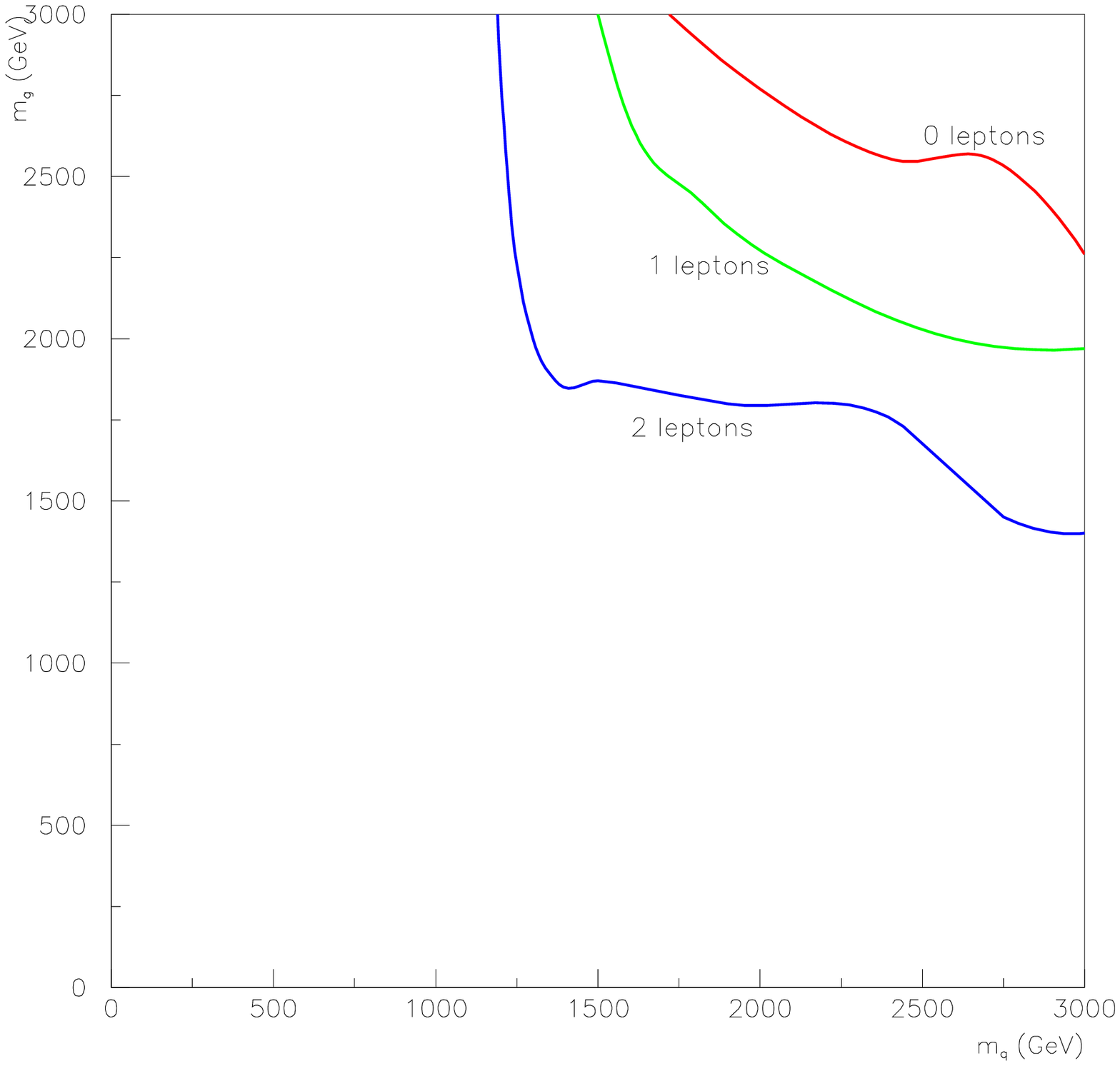}} \par}
\caption{Graph of exclusion in the plane \protect\( m_{\widetilde{q}}\, vs\, m_{\widetilde{g}}\protect \)\label{fig: exclu2}}
\end{figure}  
\begin{figure}
{\centering \resizebox*{8cm}{8cm}{\includegraphics{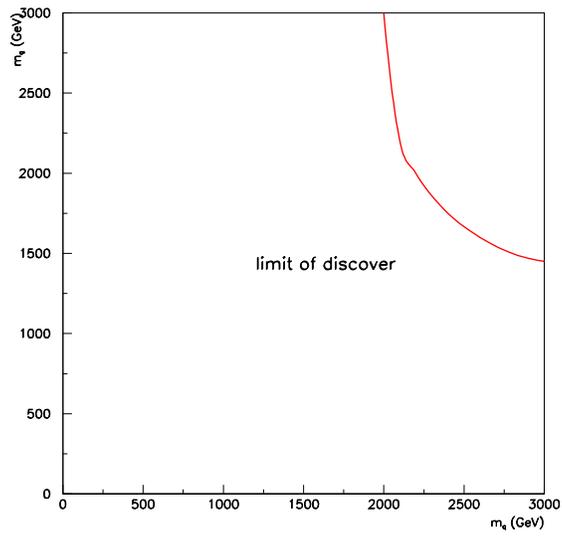}} \par}
\caption{limit of discovery of MSSM in plane \protect\( m_{\widetilde{q}}\, vs\, m_{\widetilde{g}}\protect \)
\label{fig: exclu3}}
\end{figure}
\subsubsection{Limit of discovery} 

We now will generalize this study in order to define limit of discovery for
the MSSM. The significance was calculated for high luminosity \( \int Ldt=100fb^{-1} \).
The isocurves of significance equal to 5 are given in Figure \ref{fig: exclu2} in the 
plane \( m_{\widetilde{q}} \) versus \( m_{\widetilde{g}} \),
for the set of 7 differents parameters:
\( M_{\widetilde{l}}=3000\, GeV, \) \( M_{1}=100\, GeV, \) \( M_{2}=2000\, GeV, \)
\( M_{A}=200\, GeV, \) \( \tan \beta =50, \) \( \mu =2000\, GeV, \) \( A_{3}=0\, GeV \). Each of
these isocurves represent a specific cut on the number of leptons per event.
The curves labelled \( 0l \), \( 1l \) and \( 2l \) represent significance
equal to 5 in case of each event taken into account in the calculation of
signal
and background possesses respectively \( 0 \), \( 1 \), \( 2 \) lepton(s).
Figure \ref{fig: exclu3} show the zone of discovery in plane \(
m_{\widetilde{q}}\, vs\, m_{\widetilde{g}} \)
for all the sets of parameter. There is a total of 3750 from the 17000
combinations
of parameter which wouldn't be discovered. This lead to the
limit
of discovery of about $2$ $TeV$.
 
\subsection{Conclusion}
 
We demonstrated the possibility to discover a phenomenological MSSM using an
inclusive study in the MSSM parameters space. We can note, 
at the end of our study, that we observed little
difference between MSUGRA and the pMSSM. The limit of discovery correspond to
the limit of the cross section (2.7 TeV at CMS). The only difference appears
for some points having a specific mass hierarchy. For example in the case of
close masses, the limit we expect is about 1.5 TeV.
 
\section{R parity violation }
The conservation of this quantum quantity ($R= (-1)^{3(B-L)+2S}$) allows 
to conserve baryonic and leptonic
number. Nevertheless there is no theoretical reason to impose this conservation.
We then expect several new terms to appear in the supersymmetric lagrangian depending
on a $\lambda$ coupling.
\begin{figure}[hbtp]
\begin{center}
\vspace*{0mm}
\hspace*{0mm}\resizebox{0.6\textwidth}{8cm}
                        {\includegraphics{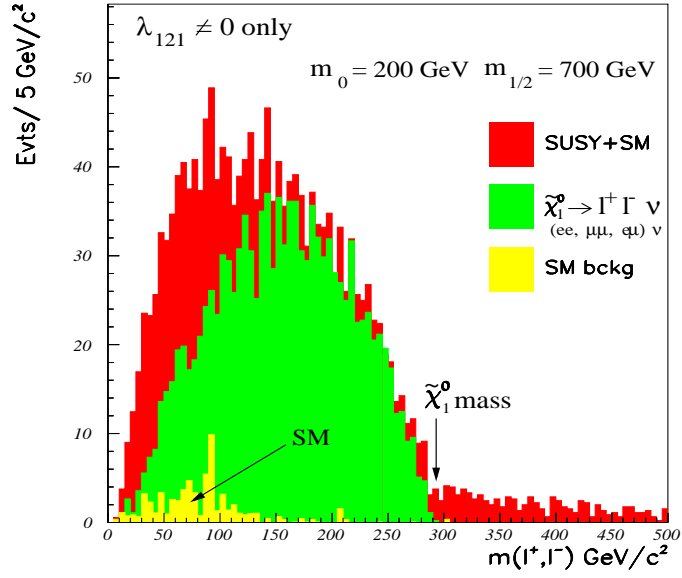}}

\vspace*{-5mm}
 \caption{Invariant dilepton mass for $\lambda_{121}>0$ }

\vspace*{-5mm}
\label{rpv1}
\end{center}
\end{figure}
Experimental constrains
lead to very small value for the factor (usually less than $\lambda<10^{-1}$) multiplying
the lagrangian terms violating the R parity. Phenomenologically for ($\lambda<10^{-6}$) we
expect the LSP to decay outside our detectors thus behaving like R-parity conserved signature.
We focused here on the intermediate situation ($10^{-6}<\lambda<10^{-2}$). The
phenomenological behaviour correspond to R-parity conserved situation with the decay of the
LSP in 3 jets or 3 leptons or 2 jets and one lepton.
Figure \ref{rpv1} illustrates the 3 leptons case where the invariant dilepton mass is plotted.
As is expected the edge of this distribution correspond to the LSP mass.
This correspond to MSUGRA model with low $m_0$ mass.

\begin{figure}[hbtp]
\begin{center}
\vspace*{0mm}
\hspace*{0mm}
\includegraphics*[width=8cm,bb=80 150 510 680]{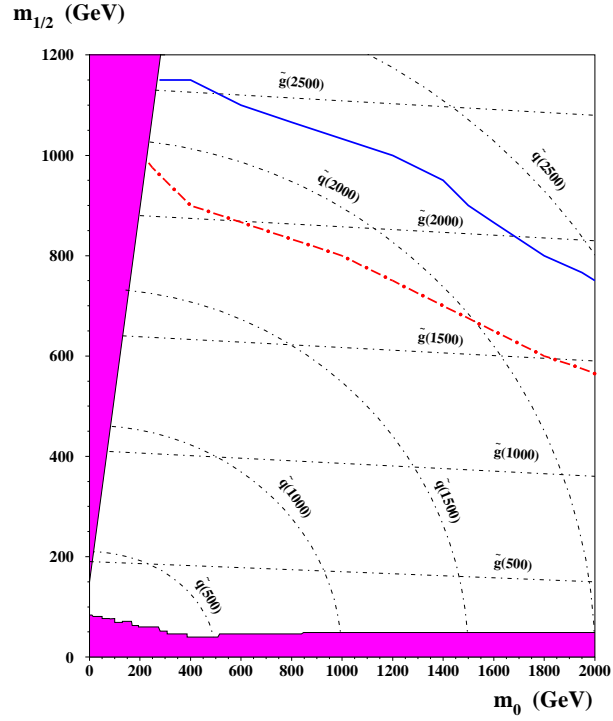}

\vspace*{-5mm}
\caption{MSUGRA mass reach for one year of low luminosity for $\lambda_{121}$ or $\lambda_{233}$ non zero}

\vspace*{-5mm}
\label{rpv3}
\end{center}
\end{figure}

The expected reach for the 3 leptons case in the MSUGRA plane is illustrated
in the following figure \ref{rpv3}.
We can note that the mass reach is roughly the same as in the R-parity
conserved one.
\begin{figure}[hbtp]
\begin{center}
\includegraphics*[height=6.5cm,bb=25 173 550 650]{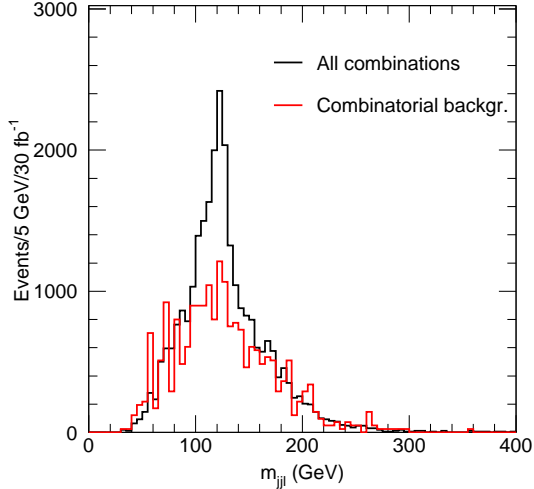} \\

 \vspace*{-5mm}
\caption{invariant dilepton and jet mass for $\lambda'$ non zero.}
\label{lambdap}
\end{center}
\end{figure}
In the following figure \ref{lambdap}, is plotted the invariant dijet and lepton invariant mass
corresponding to  $\lambda'$ non zero. We can observe a reconstructed peak at 130 GeV
matching the LSP mass.

\section{Gauge Mediated Supersymmetry Breaking}
 
In these model, SUSY breaking has its origin in a separate sector at
a relatively low scale $F_m=(10^{10}GeV)^2$(compare to MSUGRA model). An
important caracteristic
of this model is to lead to a very light gravitino mass much lower than the other
spartners. In the minimal GMSB model 6 parameter are considered:
$M_m$ messenger scale, $\Lambda=F_m/M_m$, $N_5$ number of messenger family,
$tan\beta$, $sgn\mu$, $C_{grav}$
ratio of the gravitino mass to its value if the only source of SUSY breaking is $F_m$.
For LHC experiment it is extremely interesting to evaluate the response of the detector
to the signature provided by these models as they present different phenomenology.
We expect 2 possible scenarios: \\
\begin{center}
\begin{itemize}
\setlength{\itemindent}{2mm}           
\setlength{\parindent}{2mm}            
\setlength{\itemsep}{7mm}               
\item
$\tilde{\chi_1^0} \rightarrow \tilde{G}+\gamma$ ($\tilde{\chi_1^0}$ NLSP) \\
\item
$\tilde{\l_R} \rightarrow \tilde{G}+l$ ($\tilde{\l_R}$ NLSP) \\
\end{itemize}
\end{center} 

with the following signatures:\\
\begin{center}

\begin{tabular}{|c||c|c|c|}
\hline
NSLP & $c\tau$: short & $ c\tau$: average & $c\tau$ long \\
\hline
$\tilde{\chi_1^0}$ & MSSM+$2$ $\gamma$ & $c\tau$  & MSSM  \\
 &  & measurement &  \\
 & &(ECAL,$\mu$ chamber) & \\
\hline
$\tilde{\l_R}$ & MSSM+$2$ $l$ & $c\tau$ and  & mass    \\
 & &mass  & measurement\\
 & & measurement& (TOF)\\
\hline
\end{tabular}
 
\end{center} 
 
\begin{figure}[hbtp]
\begin{center}
\vspace*{0mm}
\includegraphics*[height=8cm,bb=170 520 400 720]{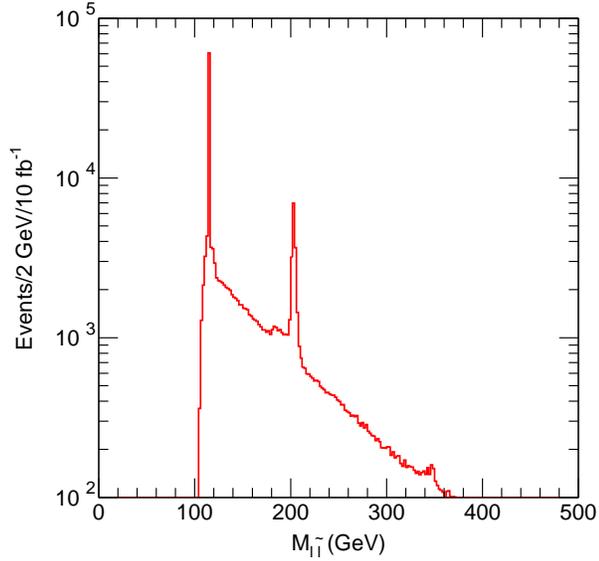} \\
\caption{Dilepton invariant mass reconstruction for slow ionizing slepton}
\label{G2B_11}
\end{center}
\end{figure}

\begin{figure}[hbtp]
\begin{center}

\includegraphics*[height=8cm,bb=200 520 420 720]{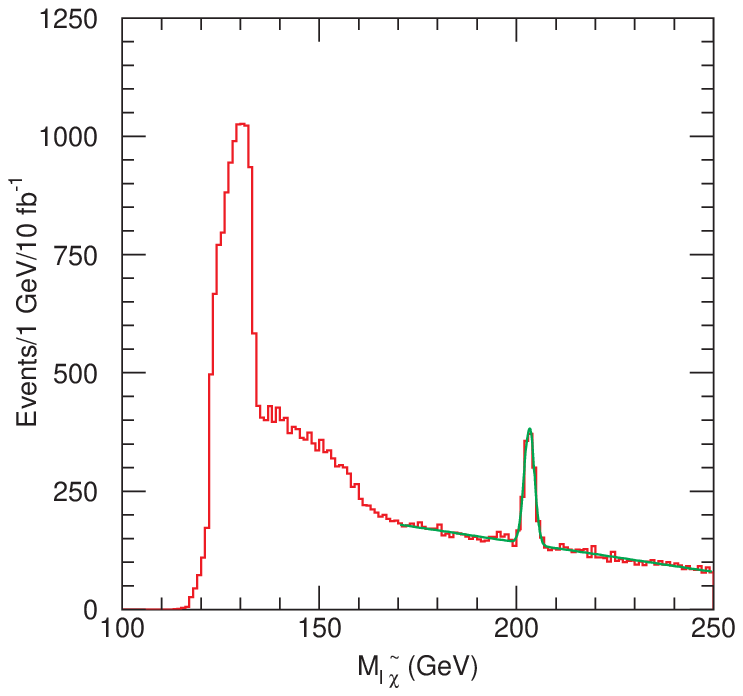} \\
\caption{Trilepton invariant mass}
\label{G2B_12}
\end{center}
\end{figure}

\begin{figure}[hbtp]
\begin{center}

\hspace*{-70mm}
\includegraphics*[height=8cm,bb=90 550 325 765]{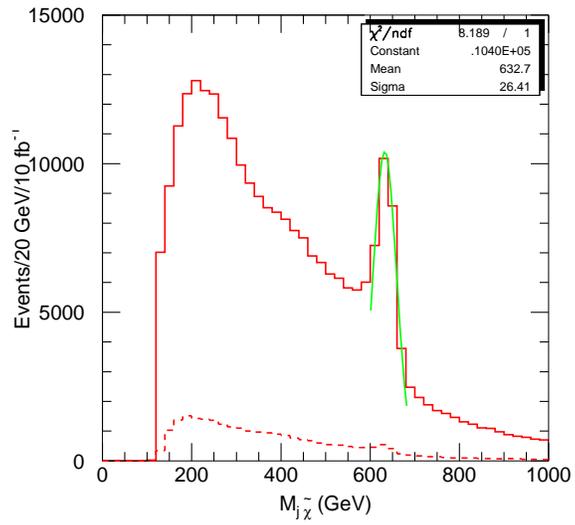} \\
\vspace*{17mm}
 \caption{Trilepton and jet invariant mass}
\label{G2B_21}
\end{center}
\end{figure}

In figure \ref{G2B_11}, we show the channel $\tilde{\chi_i^0}\rightarrow
\tilde{\l_R} l$ where we reconstruct a long living
right slepton using TOF measurement as measured in the muon chamber.
We observe 3 peaks corresponding to $\tilde{\chi_1^0}$,$\tilde{\chi_2^0}$ and $\tilde{\chi_4^0}$. Figure \ref{G2B_12} show the channel $\tilde{\l_L} \rightarrow \tilde{\chi_i^0} l \rightarrow \tilde{\l_R} l l $
: peak :$\tilde{\l_L}$, and $\tilde{\chi^{\pm}_1} \rightarrow \tilde{\nu} l \rightarrow \tilde{\l_R} \nu l l $.
Combine previous $\tilde{\chi_1^0}$ invariant mass with any of 4 hardest jets provide the reconstruction
of left squark (generated $m_{\tilde{q}}=648$ $GeV$, reconstructed 632
GeV) as can be seen in figure \ref{G2B_21}.
From this study we can deduce that the full reconstruction chain can be performed.
 
\subsection{Conclusions}
 
 The main conclusions of our studies are the following  $:$ 
 Standard Model Higgs as well MSSM Higgs should be visible
 over the complete theoretical expected range using various decay channels.
 The
SUGRA model, investigated for SUSY search, 
 would be detectable through
 an excess of events over SM expectations
up to masses m$_{\tilde{q}}$ $\sim$
m$_{\tilde{g}}$ $\sim$ 2.5 TeV
with 100 fb$^{-1}$.
 This means that the entire plausible domain of EW-SUSY
parameter space for most probable values of tan$\beta$
 can be probed.
Furthermore, the S/B ratios are $>$ 1 everywhere in the reachable domain of
parameter
space (with the appropriate cuts) thus allowing a study of kinematics of
$\tilde{q}$, $\tilde{g}$ production and obtaining information
on their masses.
 The cosmologically interesting region $\Omega h^{2} \leq$ 1, and even more
the preferred region  0.15 $\leq$ $\Omega h^{2} \leq$ 0.4,
 can be entirely  probed.
The mass reach is up to $\tilde{q},\tilde{g}= 2.5 TeV$ and the 
cosmologically interesting region is covered by CMS/ATLAS.
Sleptons can be observed up to 400 GeV, LSP up to 400 GeV.
LHC can perform exclusive studies: gluino, neutralino reconstruction
and determine MSUGRA parameters.
MSSM investigation show similarities with MSUGRA but
some specific mass hierarchy might be difficult to explore.
LHC can also strongly contribute in GMSB model and provide 
some insights for parameters determination.
R parity violated ( $\lambda <10^{-2}$) studied ( $\chi_1^0$ mass determination).
These models lead to various experimental signature with which LHC
detectors can cope.

\end{document}

%% file: econfmacros2.tex
\def\beq{\begin{equation}}
\def\eeq#1{\label{#1}\end{equation}}
\def\eeqn{\end{equation}}

\newenvironment{Eqnarray}%
   {\arraycolsep 0.14em\begin{eqnarray}}{\end{eqnarray}}
\def\beqa{\begin{Eqnarray}}
\def\eeqa#1{\label{#1}\end{Eqnarray}}
\def\eeqan{\end{Eqnarray}}

\let\bar=\overbar

\def\Dslash{\not{\hbox{\kern-4pt $D$}}}
\def\dslash{\not{\hbox{\kern-2pt $\del$}}}

\def\msb{{\bar{\ssstyle M \kern -1pt S}}}

\def\lsim{\mathrel{\raise.3ex\hbox{$<$\kern-.75em\lower1ex\hbox{$\sim$}}}}
\def\gsim{\mathrel{\raise.3ex\hbox{$>$\kern-.75em\lower1ex\hbox{$\sim$}}}}